\documentclass[12pt,letterpaper]{article}
\usepackage[utf8]{inputenc}

\usepackage{graphicx,array}
\usepackage{url}
\usepackage{color}
\usepackage{latexsym}
\usepackage{amsthm}
\usepackage{amsmath}
\usepackage{amssymb}
\usepackage{amsfonts}
\usepackage[numbers,sort&compress]{natbib}
\usepackage{bm}
\usepackage{slashed}
\usepackage{mathrsfs}
\usepackage{enumerate}
\usepackage{siunitx}
\usepackage{mdframed}
\usepackage{setspace}  
\usepackage{esvect}

\usepackage{tcolorbox}%

\usepackage{hyperref} 
\hypersetup{
    colorlinks=true,       
    linkcolor=red,          
    citecolor=blue,        
    filecolor=magenta,      
    urlcolor=blue           
}
\usepackage[all]{hypcap} 
\usepackage{multirow}
\usepackage{multicol}

\usepackage{natbib}
\newcommand{\nc}{\newcommand}

\nc{\beq}{\begin{equation}}
\nc{\eeq}{\end{equation}}
\nc{\beqa}{\begin{eqnarray}}  
\nc{\eeqa}{\end{eqnarray}}  
\nc{\bit}{\begin{itemize}}  
\nc{\eit}{\end{itemize}}  

\usepackage{floatrow}
\newfloatcommand{capbtabbox}{table}[][\FBwidth]

\usepackage{blindtext}

\title{ 
 {\bf Electrobaryonic axion:} \\
 {\bf \Large hair of neutron stars}
\author{\large Yang Bai$^{\,a}$ and Carlos Henrique de Lima$^{\,b}$}
\date{\small \it 
$^a$Department of Physics, University of Wisconsin-Madison, Madison, WI 53706, USA \\
$^b$TRIUMF,  4004 Wesbrook Mall,  Vancouver, BC V6T 2A3, Canada
}
}

\begin{document}

\maketitle

\setlength{\parskip}{0.2ex}

\begin{abstract}
Axion-like particles are predicted in many physics scenarios beyond the Standard Model (SM). Their interactions with SM particles may arise from the triangle anomaly of the associated global symmetry, along with other SM global and gauge symmetries, including anomalies with the global baryon number and electromagnetic gauge symmetries. We initiate the phenomenological study of the corresponding ``electrobaryonic axion"—a particle that couples with both the baryon chemical potential and the electromagnetic field. Neutron stars, particularly magnetars, possessing high baryon density and strong magnetic fields, can naturally develop a thin axion hair around their surface. In this study, we calculate this phenomenon, considering the effects of neutron star rotation and general relativity. For axion particles lighter than the neutron star rotation frequency, the anomalous interaction can also induce the emission of axion particles from the neutron star. In the light axion regime, this emission can significantly contribute to the neutron star cooling rate.
\end{abstract}

\thispagestyle{empty}  
\newpage   
\setcounter{page}{1}  
\section{Introduction}
\label{sec:intro}
One of the well-motivated solutions to the strong CP problem is to introduce the Peccei-Quinn (PQ) $U(1)_{\rm PQ}$ symmetry~\cite{PhysRevLett.38.1440,PhysRevD.16.1791} and dynamically relax the effective CP-violating $\theta$ angle to be negligible. After the spontaneous breaking of $U(1)_{\rm PQ}$ at a high scale $f_a$, the associated pseudo-Nambu Goldstone boson or the Quantum Chromodynamics (QCD) axion~\cite{PhysRevLett.40.223,PhysRevLett.40.279} is much lighter than the QCD scale and could account for dark matter in the universe~\cite{Preskill:1982cy,Abbott:1982af,Dine:1982ah}. Beyond the QCD axion models, axion-like particles are also ubiquitous in string theory~\cite{Svrcek:2006yi} and could serve as the smoking gun for physics beyond the Standard Model (SM).

To search for axion-like particles, physicists usually investigate their potential interactions with two photons, two gluons, or a pair of electrons and positrons~\cite{PDG}. This exploration relies on the triangular anomaly of $U(1)_a$, the global symmetry associated with the axion-like particle, either with two gauge bosons or through direct couplings to fermions. On the other hand, non-trivial triangular anomalies may also exist among currents involving more than one global symmetry, such as $U(1)_a U(1)_B [U(1)_{\rm em}]$, where $U(1)_B$ represents the baryon number symmetry and $U(1)_{\rm em}$ is the electromagnetic gauge symmetry~\cite{Metlitski:2005pr}. The corresponding axion-like particle, referred to as the ``electrobaryonic axion," couples to the baryon number density along with magnetic or electric fields. Depending on the ultraviolet (UV) physics, the axion-like particle could exhibit vanishing couplings to two gauge bosons but nonzero couplings to one gauge boson, leading to different phenomenological consequences. 

Interestingly, the neutral meson fields $(\pi, \eta, \eta')$ also exhibit electrobaryonic properties through anomalous interactions with the baryon chemical potential and magnetic field~\cite{Son:2004tq}. It has been observed that the anomalous interaction of pions in an environment characterized by high density and a strong magnetic field induces the spontaneous breaking of translation invariance, leading to the formation of the chiral soliton lattice state~\cite{Son:2007ny,Higaki:2022gnw}. 

However, unlike the neutral meson fields, the axion-like particle does not produce a similar phenomenon. This is due to its interaction being suppressed by $1/f_a$, where $f_a$ is significantly above the electroweak scale and even further above the energy scales in realistic environments, including neutron stars. Nevertheless, the electrobaryonic axion could be sought in astrophysical objects or laboratory-based experiments. 

This work explores the production of axion-like particles in finite systems with a nonzero baryon chemical potential and background magnetic field. The paper focuses on two distinct systems. Firstly, we examine a simplified one-dimensional toy model that is a foundation for understanding axion production and detection in practical experimental setups. Secondly, we investigate neutron stars, known for having stronger magnetic fields and higher chemical potentials~\cite{2010arXiv1012.3208L, Baym:2017whm}.

We demonstrate that this anomalous interaction generates a coherent solution of axions localized at the system's boundary. Additionally, we explore the effects of incorporating the time dependence of the tilted-dipole approximation of the neutron star magnetic field, revealing the potential for resonance enhancement and particle emission. Considering the impact of general relativity on the deformation of the coherent state, we discuss potential experimental approaches to probe this interaction. Our findings underscore neutron star cooling as the primary means of investigating this interaction in the regime of light axions.

The remainder of this paper is organized as follows. In Section~\ref{sec:anomalyAXION}, we briefly review the anomalous interaction of general theories with axion-like particles. Section~\ref{sec:oneD} explores the one-dimensional toy model, including time-dependent sources. Section~\ref{sec:NS} applies the anomalous interaction to neutron stars, demonstrating the sourcing of axion-like particles at the boundary of these stars. We also consider gravitational effects, discussing the potential trapping of axions produced by this mechanism. In Section~\ref{sec:pheno}, we delve into the experimental signals generated by this solution. Section~\ref{sec:disc} discusses promising directions in which this interaction can be employed to probe axion-like particles. Appendix~\ref{app:infinite} contains the time-dependent solution for the one-dimensional infinite square well, demonstrating the absence of arbitrary growth in axion fields. Appendix~\ref{app:GR} provides formulas for general relativity effects, treating the neutron star with a constant energy density.

\section{Anomalous interactions of axion}
\label{sec:anomalyAXION}

For a generic axion model, the global $U(1)_a$ symmetry could have anomalies not only with gauge symmetries but also with global symmetries. Based on the SM gauge group $SU(3)_c \times SU(2)_W \times U(1)_Y$ and the global symmetries $U(1)_B$ and $U(1)_L$ together with gravity, the potential model-dependent non-vanishing triangle anomalies (with only one $U(1)_{a}$) are \footnote{In this notation we use brackets to denote the gauge symmetries.}
\beqa
&&  U(1)_a\,[U(1)_Y]^2 \,, \quad U(1)_a\,[SU(2)_W]^2 \,, \quad U(1)_a\,[SU(3)_c]^2\,, \quad U(1)_a\,[\mbox{gravity}]^2\,, \nonumber  \\ [2pt]
&&  U(1)_a\,U(1)_B\,[U(1)_Y] \,, \quad U(1)_a\,U(1)_L\,[U(1)_Y]\,, \nonumber \\ [2pt]
&& U(1)_a\,U(1)_B^2 \,, \quad U(1)_a\,U(1)_L^2  \,, \quad U(1)_a\,U(1)_B\,U(1)_L ~. 
\label{eq:more-anomalies}
\eeqa

\begin{figure}[t!]
\centering
    \includegraphics[width=0.50\textwidth]{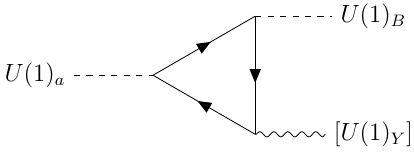}
    \caption{The Feynman diagram for the triangle anomaly of $U(1)_a\,U(1)_B\,[U(1)_Y]$ with SM fermions running in the loop. This anomaly coefficient for the electrobaryonic axion is $-c_1 - c_2$ and nonzero for the double-gauge-anomaly-free axion in \eqref{eq:non-stardard-charge} with $c_1 \neq - c_2$. }
    \label{fig:triangle}
\end{figure}

Since the coefficients of these anomalies are model-dependent, some global symmetries may have no anomalies under two SM gauge interactions but have nonzero anomalies under one SM gauge interaction and one global symmetry. A simple example of this can occur in the embedding of the SM in the SU(5) Grand Unified Theory (GUT). For generation-independent charges of SM fermions, ($q_{Q_L}$, $q_{u_R^c}$, $q_{d_R^c}$, $q_{L_L}$, $q_{e_R^c}$), under $U(1)_a$ (here, we assign charges for left-handed or charge-conjugated right-handed fermions), there are two general solutions to have vanishing anomalies of $U(1)_a\,[U(1)_Y]^2$, $U(1)_a\,[SU(2)_W]^2$, $U(1)_a\,[SU(3)_c]^2$, without introducing new fermions beyond the SM. They can be parameterized as 
\beqa
\label{eq:non-stardard-charge}
&&\mbox{\it double-gauge-anomaly-free axion:} \nonumber \\
&&\hspace{3cm}
(q_{Q_L}, q_{u_R^c}, q_{d_R^c}, q_{L_L}, q_{e_R^c}) = \left(- \dfrac{c_1}{3},-\dfrac{2\,c_1}{3}-c_2, \dfrac{4\,c_1}{3} + c_2, c_1, c_2\right) ~.
\eeqa
One global symmetry with $c_2 = - 2 c_1$ has also a vanishing anomaly for $U(1)_a\,[\mbox{gravity}]^2$. Another global symmetry with $c_1 = -3 c_2$ has zero anomalies for $U(1)_a\,[SU(5)_{\rm GUT}]^2$ with $SU(5)_{\rm GUT}$ as the GUT gauge group. The anomaly for $U(1)_a\,U(1)_B\,[U(1)_Y]$ is $-c_1 - c_2$, which is not zero for $c_1 \neq - c_2$. The triangular Feynman diagram is shown in Fig.~\ref{fig:triangle}. Interestingly, one has zero anomaly of $U(1)_a\,U(1)_{B-L}\,[U(1)_Y]$ for this class of global symmetries.

After the electroweak symmetry breaking, both anomalies of $U(1)_a\,[U(1)_Y]^2$ and $U(1)_a\,[SU(2)_W]^2$ contribute to the $U(1)_a\,[U(1)_{\rm em}]^2$ anomaly and lead to the axion coupling to two photons. A nonzero anomaly of $U(1)_a\,U(1)_B\,[U(1)_Y]$ leads to a nonzero anomaly for
\begin{align}
U(1)_a\,U(1)_B\,[U(1)_{\rm em}] \, , 
\end{align}
 which we name the Goldston boson associated with $U(1)_a$ as the ``electrobaryonic" axion (the main focus of this paper). The existence of this specific anomaly can happen in a plethora of models with nonuniversal $U(1)_{\rm PQ}$ charges, as it occurs for the DFSZ axion~\cite{DiLuzio:2020wdo}. The phenomenological consequences of this specific anomaly have not been explored before in the context of axion-like particles. In this work, we take the first step towards exploring the phenomenology of this class of anomalies.
 
 For gauge symmetries, one can use the gauge field tensors to describe their anomalous interactions with the axion-like particle associated with the spontaneous breaking of $U(1)_a$ symmetry. For the global symmetry, similar to the treatment in Refs.~\cite{Son:2004tq,Harvey:2007rd,Chakraborty:2023wgl}, we introduce two fictitious gauge fields, $\mathsf{V}_\mu$ and $\mathsf{L}_\mu$, that couple to the baryon or lepton current, respectively. The corresponding interactions for the matter field can be written as
\beqa
\mathcal{L} \supset \overline{\psi} \gamma^\mu(i \partial_\mu + q_{\rm em}\,e\,A_\mu + q_{B}\,\mathsf{V}_\mu + q_{L} \,\mathsf{L}_\mu) \psi ~,
\eeqa
where $q_{\rm em}$, $q_B$ and $q_L$ are the charges of $\psi$ under $U(1)_{\rm em}$, $U(1)_B$ and $U(1)_L$, respectively.

For the (not spontaneously broken) baryon symmetry, we replace $\mathsf{V}_\mu$ by the fermion four-current: $\mathsf{V}_\mu(t, \vec{x}) = \mu(t, \vec{x})\,u_\mu(t, \vec{x})$, with four-velocity field $u_\mu$ normalized as $u_\mu u^\mu = 1$ and $\mu$ the baryon chemical potential. In the local rest frame, one has $u_\mu = (1, \vec{0})$. In general, one has $u^\mu = \gamma(\vec{v})(1, \bm{v})$ and $u^\mu \approx (1, \bm{v})$ in the non-relativistic limit~\cite{Jaiswal:2016hex}. 

Let us focus now on the anomaly containing the baryon global symmetry. The non-conservation of the global $U(1)_a$ symmetry against $[U(1)_{\rm em}]$ and $U(1)_B$ is 
\beqa
\partial^\mu j^a_\mu =  - \frac{1}{16\pi^2}\, \left( e^2\,c_{a\gamma\gamma} \,F^{\mu\nu}\widetilde{F}_{\mu\nu} + e\,c_{a\gamma \mathsf{V}}\, \mathsf{V}^{\mu\nu} \widetilde{F}_{\mu\nu} + c_{a \mathsf{V}\mathsf{V} } \,\mathsf{V}^{\mu\nu} \widetilde{\mathsf{V}}_{\mu\nu} \right) ~,
\eeqa
with $\widetilde{F}_{\mu\nu} = \frac{1}{2} \epsilon_{\mu\nu\alpha\beta}F^{\alpha\beta}$ as the field tensor for $U(1)_{\rm em}$ and $\mathsf{V}_{\mu\nu} = \partial_\mu\mathsf{V}_{\nu} - \partial_\nu\mathsf{V}_{\mu}$ and $\widetilde{\mathsf{V}}_{\mu\nu} = \frac{1}{2} \epsilon_{\mu\nu\alpha\beta}\mathsf{V}^{\alpha\beta}$. Here, $c_{a\gamma\gamma}$, $c_{a\gamma \mathsf{V}}$ and $c_{a \mathsf{V}\mathsf{V}}$ are model-dependent anomaly factors. The derivation for the other anomalies is similar, considering the appropriate change of field strengths. 

After spontaneously breaking of $U(1)_a$, the corresponding axion or axion-like particle $a$ has the following interactions from anomaly matching
\beqa
\mathcal{L} \supset \frac{1}{8\pi^2\,f_a}\, \partial^\mu a \left( e^2\,c_{a\gamma\gamma} \,A^\nu\widetilde{F}_{\mu\nu} + e\,c_{a\gamma \mathsf{V}}\, \mathsf{V}^\nu \widetilde{F}_{\mu\nu} + c_{a \mathsf{V}\mathsf{V} } \,\mathsf{V}^\nu \widetilde{\mathsf{V}}_{\mu\nu}  \right) ~.
\eeqa
The first term generates the anomalous decay of axions into two photons. In this context, we can define the model-dependent coupling $\eta = e^{2}c_{a\gamma\gamma}/8\pi^{2}$. This coupling allows us to express the photon-axion coupling as $g_{a\gamma\gamma} = 2\eta/f_{a}$. The middle term is the one of interest, which generates the sourcing of axions at finite density. Expanding the middle term to  the first order in $\bm{v}$, we have 
\beqa
\label{eq:couplings-velocity}
\mathcal{L} \supset - \frac{e\,c_{a\gamma \mathsf{V}}}{8\pi^2\,f_a}\, \mu\,
\left[ \bm{\nabla} a \cdot \bm{B} - \partial_t a \, \bm{B}\cdot \bm{v} + (\bm{\nabla} a \times \bm{E}) \cdot \bm{v}  \right] ~.
\eeqa

This paper concentrates on the first term or the zeroth-order term in $\bm{v}$. Together with the axion mass term, the most relevant terms for our analysis are
\beqa
\label{eq:axion-simple-lag}
\mathcal{L} \supset \frac{1}{2}\partial_\mu a \partial^\mu a - \frac{1}{2}\,m_a^2\,a^2  - \frac{\kappa}{f_a}\,\mu(t, \bm{x})\,\bm{\nabla} a\cdot\bm{B}(t, \bm{x})\,  ~.
\eeqa
where we have defined a dimensionless parameter $\kappa \equiv e\,c_{a\gamma \mathsf{V}}/(8\pi^2)$. In this work, we will not consider axion self-interaction effects. This describes the system's dynamics well, provided that $a/f_{a}\ll 1$. We can then explore systems with finite boundaries to understand the effects of this anomalous interaction.

\section{One-dimensional axion source}
\label{sec:oneD}

In this section, we explore a simple one-dimensional example with both static and monochromatic time-dependent magnetic fields. The toy model is a baryon-dense slab immersed in an external magnetic field. Note that this does not replace the neutron star case, where a dipole magnetic field and an (approximately) spherical dense environment are anticipated. 
Nevertheless, in one dimension, we can already observe several features of this sourcing mechanism that can be translated to the three-dimensional case. 

To simplify the analysis, we consider a constant baryon chemical potential with a sharp boundary
\begin{align}
\mu(z) = \mu_{0} \big[H(z+z_{0})-H(z-z_{0}) \big] \, ,
\end{align}
where $H(x)$ is the Heaviside step function.

\subsection{Static magnetic field}
Let us first consider the case of a static magnetic field chosen to be in the $\hat{z}$ direction with $\bm{B} = B_z\,\hat{z}$. The equation of motion for the axion field can be derived from \eqref{eq:axion-simple-lag} as
\beqa
\label{eq:eom1}
\ddot{a} - \bm{\nabla}^2 a + m_a^2 \, a  =   \frac{\kappa}{f_a} \, \bm{\nabla} \mu(\bm{x})\cdot \bm{B} =  \frac{\kappa}{f_a} \, B_z \,\mu_0 \big[ \delta(z+z_{0}) - \delta(z-z_{0})\big] \, .
\eeqa
 The static axion profiles can be easily solved by matching the boundary conditions at $\pm z_0$. The solution is unique and has odd parity:
 \beqa
a(z) = \dfrac{\kappa\,B_z\,\mu_0}{f_a\,m_a}\begin{cases}
-\,e^{-\,m_a \, z} \sinh(m_{a}z_{0})~, & z > z_{0}~,   \vspace{3mm}\\ 
-\,e^{-m_a \, z_0} \sinh(m_{a}z)~, & |z| \leq z_{0}~, \vspace{3mm}\\ 
\,e^{m_a \, z} \sinh(m_{a}z_{0})~, & z < -z_{0} ~. 
\end{cases}
\label{eq:static-step-odd}
\eeqa

\begin{figure}[t!]
\centering
    \includegraphics[width=0.65\textwidth]{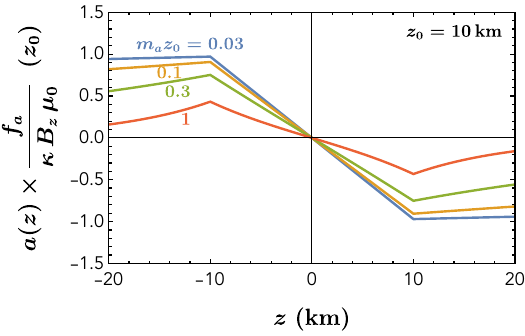}
    \caption{Examples of static axion profiles for different axion masses.
    }
    \label{fig:example1}
\end{figure}
This static profile has the characteristic behavior of being localized at the boundary. We can see that it has the characteristic form of
\begin{align}
\label{eq:one-dim-F}
 a(z)= \dfrac{\kappa\,B_z\,\mu_0}{f_a\,m_a} \,F(m_{a}z_{0},m_{a} z) \, ,
\end{align}
where we can identify a dimensionless profile function that depends on the relation between the axion mass and the size of the baryon slab. The peak value at the location of the boundary is  
\beqa
|a(z_{0})| = \dfrac{\kappa\,B_z\,\mu_0}{2\,f_a\,m_a} \left(1 - e^{-2m_{a}z_{0}} \right) \, .
\eeqa
In the heavy mass limit with $m_a z_0 \gg 1$, the peak value has the scaling of $1/m_a$. In the light mass limit, the peak value is approximately independent of $m_a$ and has a common value of $\kappa\, B_z\,\mu_0\,z_0/f_a$. In general, we have that this production mechanism prefers axions lighter than $1/z_{0}$. Some examples of the dimensionless static profiles based on \eqref{eq:static-step-odd} are shown in Fig.~\ref{fig:example1}.

\subsection{Monochromatic time-dependent magnetic field}
Given the time dependence of the neutron star magnetic field, we also include a monochromatic time dependence for the external magnetic field in the one-dimensional toy model. The magnetic field has the form $\bm{B} = B_{z}\,\cos{(\Omega \,t)}\, \hat{z}$. The equation of motion has a time-dependent source and is given by
\beqa
\label{eq:eom2}
\ddot{a} - \bm{\nabla}^2 a + m_a^2 \, a  =  \frac{\kappa}{f_a} \, B_z \, \mu_{0} \, \cos{(\Omega \,t)} \, \big[ \delta(z+z_{0}) - \delta(z-z_{0})\big] \, .
\eeqa
This equation still has a simple solution. Using an ansatz of separation of variables, $a(t, z) = \tilde{a}(z) \, \cos{(\Omega\, t)}$, the equation of motion becomes
\beqa
- \partial^2_z \tilde{a}(z) + (m_a^2 - \Omega^2) \tilde{a}(z) = \frac{\kappa}{f_a}\,B_z\,\mu_0\, \big[ \delta(z+z_{0}) - \delta(z-z_{0})\big] \, . 
\eeqa
For $m_a > \Omega$, one can define $\overline{m}_a^2 \equiv m_a^2 - \Omega^2$ and obtain the same quasi-static solution as Eq.~\eqref{eq:static-step-odd} by $m_a \rightarrow \overline{m}_a$. For $m_a < \Omega$, there is no such simple localized solution. On the other hand, one has a propagating axion field solution as
\beqa
a(t, z) = \dfrac{\kappa B_z \mu_0}{f_a}\cos{(\Omega\, t)} \, \times \,\begin{cases}
-\dfrac{\sin{(k z_0)}\cos{(k z)}}{k}~, & z > z_{0}~,   \vspace{3mm}\\ 
-\dfrac{\cos{(k z_0)}\sin{(k z)}}{k}~, & |z| \leq z_{0}~,  \vspace{3mm}\\ 
\dfrac{\sin{(k z_0)}\cos{(k z)}}{k}~, & z < - z_{0}~, 
\end{cases}
\label{eq:quasi-static}
\eeqa
with $k^2 = \Omega^2 - m_a^2 > 0$. This solution suggests that the dense slab radiates axion particles. Based on the above solution, we can also see the possibility of enhanced axion productions when the oscillation frequency matches the axion mass, $\Omega \rightarrow m_a$. With the resonance effect, one has
\beqa
a(t, z)_{\text{R}} = \dfrac{\kappa B_z \mu_0 z_{0}}{f_a}\cos{(\Omega\, t)} \, \times \,\begin{cases}
-1~, & z > z_{0}~,   \vspace{3mm}\\ 
-\dfrac{z}{z_{0}}~, & |z| \leq z_{0}~,  \vspace{3mm}\\ 
1~, & z < - z_{0}~, 
\end{cases} \qquad\qquad (\mbox{resonance:}\;\Omega = m_a)~.
\label{eq:quasi-static-limit}
\eeqa
The larger the source size, the more axions can be generated. This resonance limit is difficult to reach in realistic scenarios because the magnetic field may not have a well-behaved dipole structure with a monochromatic time dependence. Additionally, one should note that the resonance effect does not grow the axion field indefinitely. This is because the energy spectrum is continuous. In the presence of a confining potential, there can be a stronger resonance effect as one can tune the frequency of the source to match the axion discrete bound state energies of the confining potential. We explore this system, including an infinite well to mimic the effects expected from the NS gravitational well in Appendix~\ref{app:infinite}.

The formation time of these quasi-static or propagating solutions is generally short compared to the macroscopic time scales of neutron stars. The axion particles produced from the source are usually relativistic and take a short time to travel from the boundary source location to the center. 

Also, note that the axion self-interaction effects become important only when the axion field value is comparable to the decay constant. In the limit of $|a(t, z)| / f \ll 1$, which can be easily satisfied with $f_{a}^{2} \gg \kappa B_z \mu_0 z_{0}$ even in the resonance case, the axion self-interaction effects can be neglected. 

\section{Axion hair of neutron stars}\label{sec:NS}
Suppose there is an axion-like particle with a nonzero anomaly, as described in Fig.~\ref{fig:triangle}. In this case, we aim to demonstrate the existence of a thin axion hair around neutron stars, akin to the one-dimensional example in the previous section. We treat the neutron star as a spherical object with a constant energy density or baryon chemical potential and a sharp boundary. The magnetic field axis has a nonzero angle from the self-rotation axis. Similarly to the one-dimensional model, we will choose a monochromatic time dependence for the magnetic field. We first calculate the generated axion field, ignoring the curvature effects as a first approximation. We later include the deformations coming from the curved spacetime (the metric of the neutron star with a constant energy density can be found in Appendix~\ref{app:GR}). 

For simplicity, we assume that the baryon chemical potential changes only in the radial direction and can be approximated by a spherically symmetric step function. Its radial derivative can be described as
\beqa
\partial_{r}\,\mu(r) = -\mu_{0} \, \delta(r-R) \, ,
\eeqa
with $R$ as the neutron star radius. Since the source from the chemical potential change is mostly localized on the surface, the dipole approximation for the magnetic field is a good description. We denote the angle between the NS rotation axis and magnetic axis as $\alpha$ and the rotating frequency as $\Omega$ ($\sim 10^{-2}-10^{2}\,\mbox{s}^{-1}$)~\cite{Olausen:2013bpa,Kaspi:2017fwg}. With the assumption that the chemical potential changes only radially, only the radial component of the magnetic field matters and is given by
\begin{align}
B_{r}(r, \theta, \phi) &= B_{0}\frac{R^{3}}{r^{3}} \big[\sin\alpha \sin \theta \cos(\phi-\Omega\, t) + \cos\alpha \cos\theta \big] \, ,
\end{align}
where $B_{0}$ is the magnetic field strength on the NS surface of the magnetic pole.

To solve the axion equation of motion, it is convenient to express the magnetic field component in terms of the spherical harmonic functions $Y_{l}^{m}(\theta, \phi)$
\begin{align}
B_{r}(r, \theta, \phi) = B_{0}\frac{R^{3}}{r^{3}} \left(\sqrt{\frac{2\pi}{3}}\sin\alpha \big[Y_{1}^{-1}(\theta,\phi)e^{i\Omega t}-Y_{1}^{1}(\theta,\phi)e^{-i\Omega t} \big] + \sqrt{\frac{4\pi}{3}}\cos\alpha \,  Y_{1}^{0}(\theta,\phi)  \right) \, ,
\end{align}
Because the dipole magnetic field contributes only $l=1$ partial wave, the general solution for the axion field follows the same partial wave expansion as
\beqa
 a(t,r,\theta,\phi) &= R_{10}(t,r) \sqrt{\dfrac{4\pi}{3}}Y_{1}^{0}(\theta, \phi) + R_{1-1}(t,r) \sqrt{\dfrac{2\pi}{3}} Y_{1}^{-1}(\theta, \phi) - R_{1+1}(t,r) \sqrt{\dfrac{2\pi}{3}} Y_{1}^{1}(\theta, \phi) \, . \, \, \, \, \, 
 \label{eq:partial-wave-expansion}
\eeqa
%

\subsection{Flat spacetime}
\label{sec:flat}
We first ignore the curved spacetime effects from the neutron star's gravitational mass. Substituting \eqref{eq:partial-wave-expansion} into the axion equation of motion in the Minkowski spacetime, the radial wave functions satisfy
\beqa
\ddot{R}_{10} +\left(\widehat{L}_{1} + m_{a}^{2} \right)R_{10}  &=& -\frac{\kappa\mu_{0}B_{0}}{f_{a}} \, \cos{\alpha}\,\delta(r-R) \, , \\
\ddot{R}_{1\mp 1} +\left(\widehat{L}_{1} + m_{a}^{2} \right)R_{1\mp 1}  &=& - \frac{\kappa\mu_{0}B_{0}}{f_{a}} \, \sin{\alpha} \, e^{\pm i\Omega t}\,\delta(r-R) \, ,
\eeqa
where the spherical Bessel operator is defined to be $\widehat{L}_{1}= - \dfrac{1}{r^{2}} \dfrac{\partial}{\partial r}\left( r^{2} \dfrac{\partial}{\partial r} \right) + \dfrac{2}{r^{2}}$ for $l = 1$.

One can see that the structure of the equation is similar to the one-dimensional case. The interesting feature now is that we have both a time-independent source for the $R_{10}$ mode and a time-dependent source for the $R_{1\mp1}$ mode simultaneously. 

\begin{figure}[t!]
\centering
      \resizebox{0.49\linewidth}{!}{ \includegraphics{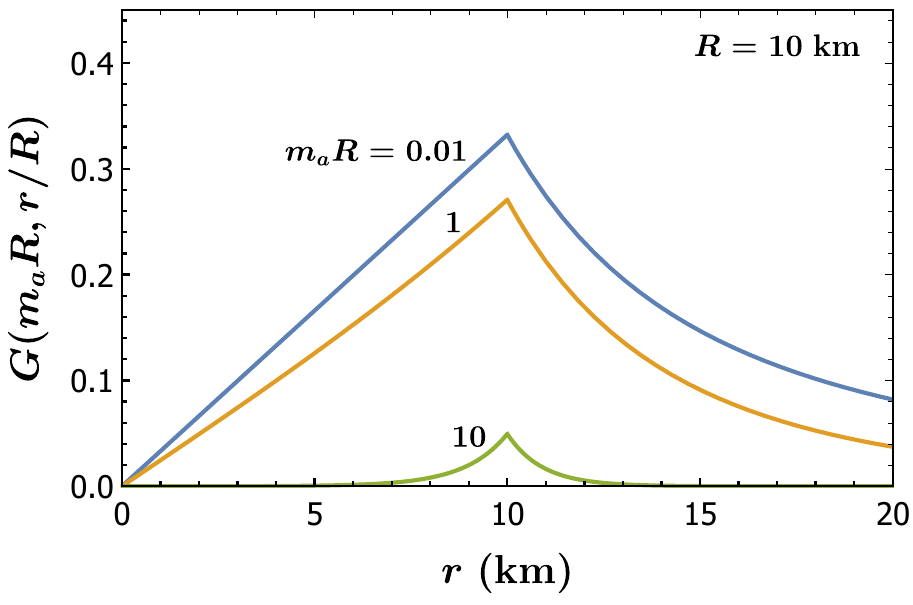}}
            \resizebox{0.49\linewidth}{!}{ \includegraphics{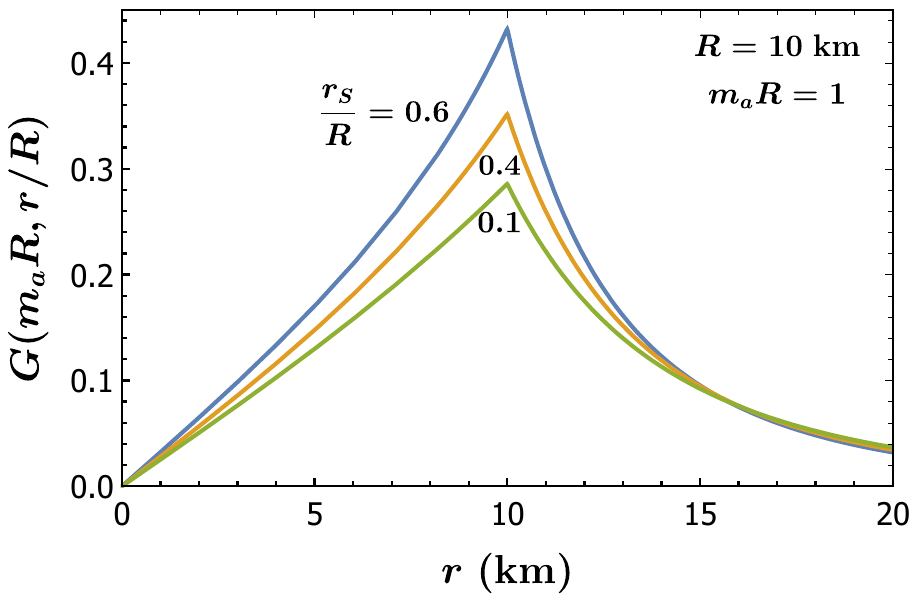}}
    \caption{\label{fig:examplespherical} Left panel: the radial profiles for the static solution in \eqref{eq:R10-static} with different values of $m_{a}R$ in flat spacetime. Right panel: the same as the left panel but including the curved spacetime effects with different neutron star masses or Schwarzschild radii $r_{\rm S} \equiv 2 G_N M$ and a fixed value of $m_{a}R = 1$. The expected range of Schwarzschild radii ranges from $0.15\,R$ to $0.65\,R$~\cite{Demorest:2010bx}, which indicates that the deformation from curved spacetime effects is subdominant.}
\end{figure}

For the mode $R_{10}$, the static solution is
\beqa \label{eq:static-sphere}
R_{10} = -\dfrac{\kappa\,B_0 \, \mu_0 \, R\,\cos{\alpha} }{f_a}\begin{cases}
\,m_a R \,  k_{1}(m_{a}R)\,  \, i_{1}(m_{a}r) \, , &\quad   r \leq R \, ,   \vspace{3mm}\\ 
\, m_a R \, i_{1}(m_{a}R)  \, k_{1}(m_{a}r) \, , &\quad r > R \, ,
\end{cases}
\eeqa
where the modified spherical Bessel functions $i_{1}(x)= (x \cosh{x} - \sinh{x})/x^2$ and $k_{1}(x) = e^{-x}(x+1)/x^2$. Note that the static solution of $R_{10}$ has a similar form as the solution of the one-dimensional case [see Eq.~\eqref{eq:one-dim-F}] and is given by
\begin{align}
\label{eq:R10-static}
R_{10} = -\dfrac{\kappa\,B_0 \, \mu_0 \, R\,\cos\alpha   }{f_a} \, G(m_{a}R,m_{a}r) \, ,   
\end{align}
where we have a dimensionless profile function $G(m_{a}R,m_{a}r)$.  Its value can be seen in the left panel of Fig.~\ref{fig:examplespherical} with different values of $m_a R$. In the right panel, we incorporate the general relativity effects generated by the neutron star on the profile function, which we discuss further in Section~\ref{sec:curved}.

Note that the function $G(m_a R, m_{a}R)$ has the asymptotic behavior of $1/(2 m_a R)$ for $m_a R \rightarrow \infty$ and $1/3$ for $m_a R \rightarrow 0$. Like the one-dimensional case, the axion peak amplitude at $r=R$ is suppressed by $1/m_a$ in the heavy axion mass limit and is independent of the axion mass in the light axion mass limit. 

For the other two modes, $R_{11}$ and $R_{1-1}$, when $\Omega=0$, one has the same static solution as $R_{10}$ by making the change of $\cos\alpha \rightarrow \sin \alpha$ in \eqref{eq:R10-static}.   
When $\Omega \neq 0$, there exist simple time-dependent solutions using the ansatz of separation of variables, $R_{1\mp 1}(t,r)=e^{\pm i\Omega t} \widetilde{R}_{1\mp 1}(r)$. 
For $m_a > \Omega$, one can define $\overline{m}_a^2 \equiv m_a^2 - \Omega^2$ and obtain the same solution for $\widetilde{R}_{1\mp 1}(r)$ as Eq.~\eqref{eq:static-sphere} by $m_a \rightarrow \overline{m}_a$. For $m_a < \Omega$, the propagating solution is
\begin{align}\label{eq:propagating}
R_{1\mp1} = -\dfrac{\kappa\,B_0\, \mu_0  \,R\, \sin\alpha}{f_a}\,e^{\pm i\Omega t} \begin{cases}
\,k R \, j_{1}(k R)\,y_{1}(k r) \, , &  \quad r \leq R~ ~,   \vspace{3mm}\\ 
 \,k R\,y_{1}(k R) \, j_{1}(k r) \, , & \quad r > R~  ~,
\end{cases}
\end{align}
where $j_{1}(x) = \sin{x}/x^2 - \cos{x}/x$ and $y_{1}(x) = - \cos{x}/x^2 - \sin{x}/x $ are the spherical Bessel functions and $k = \sqrt{\Omega^2 - m_a^2}$.

Similar to the one-dimensional case, there is a resonance enhancement effect when $\Omega = m_{a}$ with
\begin{align}
R_{1 \mp1}^{\rm R} = -\dfrac{\kappa\,B_0\, \mu_0  \, R\,  \sin\alpha }{f_a}e^{\pm i\Omega t} \begin{cases}
\,  \dfrac{r}{3\,R} \, , &  r \leq R~ ,  \vspace{3mm}\\ 
  \dfrac{R^{2}}{3\,r^{2}}\, , & r > R~  ~,
\end{cases} \qquad\qquad (\mbox{resonance:}\;\Omega = m_a)~.
\label{eq:R1mp1-resonance}
\end{align}
The enhancement is still proportional to the neutron star's radius with a geometric factor of $1/3$.

\subsection{Curved spacetime}
\label{sec:curved}
So far, we have ignored the NS gravitational potential or the effects of general relativity. Naively, the constant source on the surface of NS can continuously generate axion particles. If the axion particle speed is lower than the escaping speed inside the NS gravitational potential well, the generated axion particles will be ``stuck"  around the NS. Given enough time, the accumulated axion particles or field values could be tremendous till the axion self-interaction effects sit in. However, we have found that this naive picture is invalid as the wave or quantum effects of the axion field suppress the arbitrary growth of the axion field surrounding the neutron star, in contrast to the results in Ref.~\cite{Noordhuis:2023wid} with a large ``axion cloud". The only possibility of linear growth occurs when the frequency of the source matches one of the bound state energies. Outside of this regime, the expected solution has a form similar to the one derived in the last section. In Appendix~\ref{app:infinite}, we use a one-dimensional model with an infinite square well to analytically demonstrate that the generated axion field outside the resonance eventually matches the quasi-static solution.

In this section, we include the general relativity effects using a spherically symmetrical Schwarzschild-like spacetime~\footnote{One could also include the NS rotation effects using Kerr-like non-spherically symmetric metric. The NS rotation angular momentum is not that large to include the non-spherical spacetime.}.
To simplify our discussion and have a more analytic diagnosis, we use a constant energy density to model the inside of an NS. The solution to the Einstein equations is discussed in the classical paper by Tolman~\cite{Tolman:1939jz}. We refer the reader to Appendix~\ref{app:GR} for detailed derivations and formulas. After partial-wave expansion, the equation of motion for the $R_{1\mp1}$ mode is 
\beqa
&& g_{tt}^{-1}\,\partial_t^2 \,R_{1\mp1}\, + \, |g|^{-1/2}\, \partial_r \left( |g|^{1/2}\,g_{rr}^{-1}\, \partial_r R_{1\mp1} \right) +  \left(\frac{2}{r^2} + m_a^2 \right)\,R_{1\mp1} \nonumber \\
&& \hspace{9cm} = - \frac{\kappa\,\mu_{0}\,B_{0}}{f_{a}} \, \sin{\alpha} \, e^{\pm i\Omega t}\,\delta(r-R)
~,
\label{eq:R1mp-curved}
\eeqa
with the formulas for $g_{tt}$, $g_{rr}$ and $|g|$ in Appendix~\ref{app:GR}. The equation of motion for $R_{10}$ is the same by setting $\Omega=0$ and $\sin{\alpha}\rightarrow \cos{\alpha}$.

The results of the static or quasi-static solutions are similar to the flat-spacetime case. Defining the ``Schwarzschild radius" for the neutron star $r_{\rm S} \equiv 2 G_N M$ with $G_N$ as the Newton constant and $M$ as the neutron star mass, the profiles of the static solutions of $R_{10}$ are shown in the right panel of Fig.~\ref{fig:examplespherical}. One can see that the curved spacetime does not change the static profiles significantly.  

Like the one-dimensional case, the time-dependent source could generate a resonance effect to enhance the axion hair around the neutron star. The resonance effects happen when the rotating frequency, $\Omega$, matches the bound state energy levels. For the toy infinite square well in Appendix~\ref{app:infinite}, the axion bound state energy levels are simply $E_n \equiv \sqrt{m_a^2 + n^2\pi^2/(4 L^2)}$ with $2L$ as the potential well length. The resonance enhancement factor is $1/\epsilon$ with $\epsilon \equiv (\Omega - E_{n_R})/\Omega$ when $\Omega$ is close to one of bound state energy $E_{n_{\rm R}}$. 

\begin{figure}[t!]
\centering
    \includegraphics[width=0.60\textwidth]{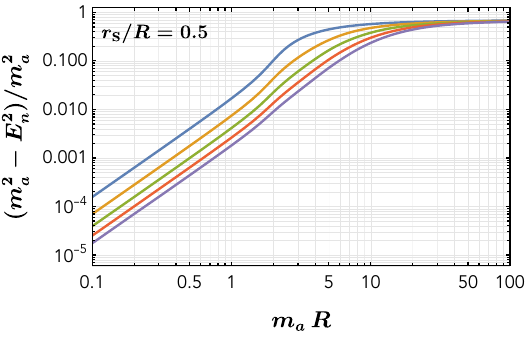}
    \caption{The bound state energy levels for different axion masses. The neutron star mass is fixed to have $r_{\rm S} \equiv 2 G_N M = 0.5\,R$. The first five states are shown here, with the top one as the ground state. }
    \label{fig:eigenvalues}
\end{figure}

One can also calculate its relativistic bound state energy levels for the axion field in the curved spacetime. We define the following ``Hamiltonian-square" operator 
\beqa
\widehat{H^2} \equiv g_{tt}\,|g|^{-1/2}\, \partial_r \left( |g|^{1/2}\,g_{rr}^{-1}\, \partial_r \right) +  g_{tt}\,\left(\frac{2}{r^2} + m_a^2 \right) ~. 
\eeqa
The bound state energy levels are $\widehat{H^2} \widetilde{R}_{1\mp1}(r) = E^2 \widetilde{R}_{1\mp1}(r)$ with Dirichlet boundary conditions $\widetilde{R}_{1\mp1}(r=0)=\widetilde{R}_{1\mp1}(r=+\infty)$. In Fig.~\ref{fig:eigenvalues}, we numerically solve the eigenvalue problem and obtain the first five bound state energies for different axion masses, where the neutron star mass and radius relation is fixed to have $r_{\rm S} /R = 0.5$. Note that in the light axion mass limit with $m_a R \ll 1$, the de Broglie wavelength of the axion is much larger than the neutron star radius. Based on the ``gravitational atom" or hydrogen-like picture, the analytic formula
\begin{align}
    E_i^2 = m_a^2\left[ 1- \dfrac{1}{4}\left(\dfrac{r_{\rm S}\,m_{a}}{i+1} \right)^{2}\right] \, ,
\end{align}
with $i=1,2,\cdots$, agrees with the numerical answers in Fig.~\ref{fig:eigenvalues} (see also Ref.~\cite{Day:2019bbh}). In the heavy axion mass limit with $m_a R \gg 1$, the axion wavefunction is very localized around the center of the neutron star such that only a small fraction of the total neutron star mass provides the gravitational potential on the axion field. The ground state energy reaches a $m_a$-independent constant ratio: $(m_a^2 - E_1^2)/m_a^2 = c_1$. The excited states have a simple scaling $E_i^2 - E_1^2 = c_2\,i\,m_a/r_{\rm S}$ with $i=2,3\cdots$. Here, both constants $c_1$ and $c_2$ depend on $r_{\rm S}/R$. The behaviors of various bound state energy levels can be found in Fig.~\ref{fig:eigenvalues}.

The resonance effects can be derived (at least algebraically) using the Hilbert space of the bound states. The bound solutions of Eq.~\eqref{eq:R1mp-curved} can be expanded into the basis eigenfunctions of $\widehat{H^2}$ as
\beqa
 R_{1\mp1}(t, r) = \sum_i \, d_i \, \widetilde{R}^i_{1\mp1}(r)\,e^{\pm i \Omega t} ~,
\eeqa
with the coefficients $d_i$ to be determined and $\widetilde{R}^i_{1\mp1}(r)$ the eigenfunctions. The equation of motion on this basis is
\beqa
\sum_i d_i\,( E_i^2 - \Omega^2 ) \widetilde{R}^i_{1\mp1}(r) =  - \frac{\kappa\,\mu_{0}\,B_{0}}{f_{a}} \, \sin{\alpha} \, g_{tt}(R)\,\delta(r-R) ~,
\eeqa
where $g_{tt}(R) = 1 - r_{\rm S}/R$. Expanding the delta function on this basis as $\delta(r-R) = \sum_i e_i \widetilde{R}^i_{1\mp1}(r)$, one can write
\beqa
 d_i = - \frac{\kappa\,\mu_{0}\,B_{0}}{f_{a}} \, \sin{\alpha} \, g_{tt}(R) \, \frac{e_i}{E_i^2 - \Omega^2} ~.
\eeqa
When $\Omega$ matches to one of the resonance energy levels, $\epsilon \equiv (\Omega - E_{n_{\rm R}})/\Omega \ll 1$, the axion field value is enhanced with the resonance-enhanced mode coefficient as
\beqa
 d_{i_{\rm R}} \approx \frac{\kappa\,\mu_{0}\,B_{0}}{f_{a}} \, \sin{\alpha} \, g_{tt}(R) \frac{e_i}{2\,\epsilon\,\Omega^2} ~,
\eeqa
with $\epsilon \ll 1$. The resonance-enhanced quasi-static axion field is then dominated by one single mode with the form as
\beqa
R^{\rm R}_{1\mp1}(t, r) \approx d_{i_{\rm R}} \, \widetilde{R}^{i_{\rm R}}_{1\mp1}(r)\,e^{\pm i \Omega t} ~.
\eeqa
We identify the eigenfunctions $\widetilde{R}^i_{1\mp1}(r) = m_a^{-1/2} R_{n1}(r)$ with $n= i + 1$ and $R_{n1}(r)$ as the orthonormal Hydrogen radial wave function.

For the neutron stars that we are interested in, the rotating frequencies are around $10^{-2}-10^2\,\mbox{s}^{-1}$, which means that the resonance effects happen for $m_a \approx \Omega$. In this regime, we have $m_a \,R \approx \Omega \, R\ll 1$, and the gravitational atom describes the bound system well. In this limit, we can identify $r_{\rm S} m_a/2 = G_N M m_a \rightarrow \alpha$ as the fine-structure constant and $a = 2/(m_a^2 \, r_{\rm S})$ as the Bohr radius. Since our system has $l=1$ angular momentum, the principle quantum number $n$ is related to the energy level index ``$i$" as $n = i + 1$. 

 If the resonance effect happens when $\Omega$ is close to the 2p state with $i=1$ and $n=2$, one can write the axion profile approximately as
\beqa \label{eq:resA}
R^{\rm R}_{1\mp1}(t, r) \approx \,\frac{\kappa\,\mu_{0}\,B_{0}\,\sin{\alpha} \, g_{tt}(R)}{f_{a}} \,  \frac{m_a^{10}\,R^3\,r_{\rm S}^{5}}{1536\,\epsilon\,\Omega^2} \,r\,e^{-r\,r_{\rm S}\,m_a^2/4}\,e^{\pm i \Omega t} ~,
\eeqa
using the Hydrogen-atom radius wave function $R_{21}(r) = \frac{1}{\sqrt{24}}\,a^{-3/2}\,\frac{r}{a}\,e^{-r/2a}$. More generally, we can define a resonance ``$Q$" factor for the bound state energy $i$ as
\begin{align}
Q_{i} \equiv \dfrac{\Omega^{2}}{E_{i}^{2}-\Omega^{2}} ~.
\end{align}
Based on the example for the 2p state in Eq.~\eqref{eq:resA}, we expect the general solution to scale with $Q_i$ when the frequency $\Omega$ is close to a specific bound state energy. 

The radial profile, in general, can be written as 
\begin{align}
\label{eq:curved-res}
|R_{1\mp1}| \approx \dfrac{\kappa\,B_0 \, \mu_0 \, R\,\sin\alpha   }{f_a} \, g_{tt}(R)  \, Q_i \,  G_{\text{NS}}(m_{a}R,m_{a}r) \, ,   
\end{align}
where one has the enhancement factor $Q_i$, and a geometric profile factor that includes the spacetime effects $G_{\text{NS}}(m_{a}R,r/R)$, which play an important role and compensate the resonance enhancement effect. For the 2p state example in  Eq.~\eqref{eq:resA} and on the surface of the NS, one can identify
\beqa
G_{\text{NS}}(m_{a}R, m_{a}R) = \frac{m_{a}^{8} \, R^{3}\, r_{\rm S}^{5}}{768} = \frac{m_a^8\,R^8}{24576} ~,
\eeqa
where we choose $r_{\rm S}= 0.5\,R$ for the second equality. One can see a significant reduction of $(m_a R)^8$ from the geometric factor in the $m_a R \ll 1$ limit. This suppression becomes stronger when exciting higher energy levels. 

For the NS radius and rotation frequencies considered here, one has $m_{a}R \approx  3.3 \times [10^{-7}, 10^{-3}]$. In the best-case scenario, one needs a resonant enhancement factor of order $Q_i = O(10^{25})$ to counter the geometric suppression factor. This restricts the resonance effect within a tiny mass window to match (almost perfectly) different NS rotation frequencies. Even when this happens, one must consider other effects that affect the resonance effect beyond the simplified treatment here: a spherically symmetric NS with a constant rotation frequency. 

\section{Elusive nature of the axion hair}\label{sec:pheno}

So far, we have shown the axion profiles generated for the one-dimensional slab and on the surface of a neutron star. For both cases, the (quasi-)static axion field profile depends on the model parameters and neutron star properties via the combination of $B_{0}\,\mu_{0}/f_{a}$ with the possible neutron star size $R$ enhancement when $\Omega \approx m_a$ [see \eqref{eq:R1mp1-resonance}]. For a generic axion model with a nonzero coupling to two photons, various astrophysical and lab-based constraints require a large value $f_{a}/\eta>10^{11}\,\text{GeV}$~\cite{PDG}. In the ultra-light axion mass regime, which is the focus of our interest, the most stringent constraints originate from star clusters~\cite{Dessert:2020lil}, supernova $\gamma$-rays conversion~\cite{Meyer:2020vzy}, and X-ray conversion in galaxy clusters~\cite{Wouters:2013hua,Marsh:2017yvc,Reynolds:2019uqt}.  The axion hair is thus highly suppressed even when considering a magnetar with a large magnetic field and baryon chemical potential. On the other hand, for the particular electrobaryonic axion without an anomalous coupling to two photons or two gluons, the current experimental constraints on the decay constant $f_a$ could be much weaker. One may consider a low value of $f_a$ as low as the TeV scale. Realizing an axion model without the two-gluon and two-photon couplings can be challenging due to the kinetic and mass mixings between the axion and neutral meson fields in the low-energy chiral Lagrangian. Nevertheless, it is possible to adjust model parameters to suppress the axion's two-gluon and two-photon couplings~\cite{Bauer:2020jbp}.

In this section, we estimate some basic properties of axion hair around the neutron star. We compare the energy of axion hair with the average dark matter energy density in the universe. We also estimate the axion emission power from a neutron star. Furthermore, we discuss how the axion hair back-reacts to the neutron star's magnetic and electric field values. Both effects are interesting but challenging to probe in a realistic setup because of the $1/f_a^2$ suppression factor.  

\subsection{Thin axion hair}

The axion hair around a neutron star is generally very thin, although its number density could be higher than the ambient axion dark matter number density. In the non-relativistic limit, the axion number density is given by $n_{a}= m_{a} |a|^{2}/2$, from which the total number of axion-like particles in axion hair around the neutron star is estimated from a volume integration using the static solution in \eqref{eq:static-sphere}. The total number of axion-like particles in the hair is
\begin{align} \label{eq:NdNS}
    N_{a}^{\rm NS-hair} \approx  (5\times 10^{41}) \, \kappa^{2} \left(\dfrac{10^{11} \text{GeV}}{f_{a}} \right)^{2} \left( \dfrac{10^{-11}\text{eV}}{m_{a}} \right)^{2} \left( \dfrac{B_{0}}{10^{15}\,\text{G}}\right)^{2} \left(\dfrac{\mu_{0}}{10^{-3}\,\text{GeV}} \right)^{2} \left( \dfrac{R}{10\,\text{km}}\right)^{2} \,\mathcal{G}(m_a R)  \, ,
\end{align}
with the geometric factor from integration
\begin{align}
   \mathcal{G}(\hat{R})  =\frac{\pi\, e^{-2 \hat{R}} \left[e^{2 \hat{R}}
   \left(\hat{R}^2-3\right)+(\hat{R}+1)\,[\hat{R}\,(2 \hat{R}+3)+3]\right]}{3\,
  \hat{R}^2} \, ,
\end{align}
where we defined the dimensionless quantity $\hat{R} \equiv m_a R$. One has $\mathcal{G}(\hat{R}) \approx 8\pi\,\hat{R}^3/45$ in the limit of $\hat{R} \rightarrow 0$ and $\pi/3$ in the limit of $\hat{R} \rightarrow \infty$. The geometric factor reaches a maximum value when $\hat{R} \approx 1$ or when $m_a \approx 10^{-11}$~eV for a neutron star with $R \approx 10$~km. The total effective mass from this solution is given by 
\begin{align} 
    M_{a}^{\rm NS-hair} \approx  \left(10^{-2}\,\mbox{g}\right) \, \kappa^{2} \left(\dfrac{10^{11} \text{GeV}}{f_{a}} \right)^{2} \left( \dfrac{10^{-11}\text{eV}}{m_{a}} \right) \left( \dfrac{B_{0}}{10^{15}\,\text{G}}\right)^{2} \left(\dfrac{\mu_{0}}{10^{-3}\,\text{GeV}} \right)^{2} \left( \dfrac{R}{10\,\text{km}}\right)^{2} \,\mathcal{G}(m_a R)  \, , \nonumber 
\end{align}
which is negligible compared to the neutron star mass. On the other hand, the number of axion particles in the hair could be larger than the one from ambient dark matter. Using the local dark matter energy density of $0.4\,\text{GeV}/\text{cm}^3$ in the solar system, the total number of axion particles around the neutron star volume is roughly $(2\times 10^{38})\times (10^{-11}\,\mbox{eV}/m_a)$, which is smaller than the number in the axion hair. Because of the scaling of $N_{a}^{\rm NS-hair} \propto m_a$ in the limit of $m_a R \ll 1$, the axion particle number in the hair is smaller than the ambient axion number for $m_a < 10^{-12}$~eV.

\subsection{Axion emission from a neutron star}
Considering the rotation of neutron stars, axion-like particles can also be emitted from a neutron star when $m_{a}<\Omega$ [see the propagating solutions in \eqref{eq:propagating}]. Matching to the amplitude of a spherical propagating wave with $\frac{A}{r}\,\cos[k\,r - \Omega\,t]$ with $k^2 = \Omega^2 - m_a^2$, one has the amplitude
\begin{align}
|A| = \dfrac{\kappa\,B_0\, \mu_0 \, R^2 \, \sin\alpha \, |y_{1}(k \, R)|}{f_a} \, .
\end{align}
Note that in the limit of $m_a < \Omega \ll 1/R$ one has $|y_{1}(k \, R)| \approx 1/(k^2 R^2)$ and $|A| \propto 1/k^2$. In the opposite limit with $k \, R \gg 1$, $|y_{1}(k\, R)|\approx 1/2$ and $|A| \propto R^2$. 

The emitted axion power from a neutron star is given by
\beqa
P &=& 4\pi\,\Omega\, k\,|A|^2 = 4\pi\,\Omega\,k\,\dfrac{\kappa^2\,B_0^2\, \mu_0^2 \, R^4 \, \sin^2\alpha \, |y_{1}(k \, R)|^2}{f_a^2}  \overset{kR \ll 1}{\approx\joinrel\approx\joinrel\approx} \frac{4\pi\,\kappa^2\,B_0^2\, \mu_0^2 \,\Omega\,\sin^2\alpha}{f_a^2\,k^3} 
\label{eq:radiation-power}
\\ [2pt]
&=& (3 \times 10^{29} \, \mbox{erg/s}) \,\kappa^{2}\, \sin^2\alpha\, \left(\dfrac{10^{11} \text{GeV}}{f_{a}} \right)^{2} \left( \dfrac{100\,\mbox{s}^{-1}}{\Omega} \right)^{2} \left( \dfrac{B_{0}}{10^{15}\,\text{G}}\right)^{2} \left(\dfrac{\mu_{0}}{10^{-3}\,\text{GeV}} \right)^{2} ~. \nonumber 
\eeqa
Here, we have used $k  \approx \Omega $ in the $\Omega \gg m_a$ limit.

The emitted axions carry away energy from both the magnetic field and the baryon chemical potential, leading to the cooling down of the star. Compared to the neutron star cooling rate $\sim 10^{32}\,\mbox{erg/s}$ from neutrino emission~\cite{Yakovlev:2004iq,Page:2004fy}, the cooling rate from emitting axion particles can be significant. For $m_a < \Omega \lesssim 100\,\mbox{s}^{-1} \approx 7\times 10^{-14}\,\mbox{eV}$, the decay constant has the following lower limit:
\beqa
\qquad \qquad f_a/\kappa \gtrsim 5\times 10^9\,\mbox{GeV}\,,  \qquad \mbox{for} \quad m_a \lesssim  7\times 10^{-14}\,\mbox{eV} ~.
\eeqa
As a first approximation, this limit does not account for the back-reaction to $B_{0}$ and $\mu_{0}$. For a heavier axion mass above the NS rotation frequency, the above cooling-rate constraint does not apply. One may need to study the axion productions from internal baryon motions inside the NS, using the sub-leading terms in velocity of \eqref{eq:couplings-velocity}.

\subsection{Back-reacted B and E fields from the axion hair}
The sourcing of axions described in this work can be understood as a coherent state of the axion-like field. They are not free axions and will not decay as such. Nevertheless, the solution sources an electric field in the direction of the magnetic field, which could be understood microscopically as the axion conversion to photon. This conversion will include another suppression factor of $1/f_{a}$. Based on the Lagrangian $\mathcal{L}= - \frac{1}{4}F_{\mu\nu}F^{\mu\nu} - g_{a\gamma\gamma}\,a\,\frac{1}{4} F_{\mu\nu}\widetilde{F}^{\mu\nu} -  g_{a\gamma V}\,a\, V_{\mu\nu}\,\widetilde{F}^{\mu\nu}$ with $g_{a\gamma\gamma} \equiv e^2\,c_{a\gamma\gamma}/(4\pi^2 f_a)$ and $g_{a\gamma V} = \kappa/(2\,f_a)$, the equations of motion of axion, $\bm{E}$ and $\bm{B}$ fields are
\beqa
\ddot{a} - \bm{\nabla}^2 a + m_a^2 \, a  &=& -g_{a\gamma \gamma} \bm{E}\cdot\bm{B}\, + \, \dfrac{\kappa}{f_{a}} \bm{B} \cdot \bm{\nabla}\mu  \, , \\ [2pt]
\bm{\nabla} \cdot \bm{E} & =& -\,g_{a \gamma \gamma} \bm{B} \cdot \bm{\nabla} a \, ,
\label{eq:E-axion}
\\  [2pt]
\bm{\nabla} \times \bm{B} & =&\frac{\partial \bm{E}}{\partial t}\,-\,g_{a \gamma \gamma}\,(\bm{E} \times \bm{\nabla} a-\bm{B}\, \dot{a}) \, + \, \dfrac{\kappa}{f_{a}} \bm{\nabla} a \times \bm{\nabla}\mu \, .
\label{eq:nablaB}
\eeqa

The interaction explored in this work generates a new magnetic field source, which can be the dominant effect for double-gauge-anomaly-free axions with $g_{a\gamma\gamma}=0$ models. One can explore the effect of this term at the leading order in $\kappa$ by considering only the additional source term from the axion hair. The equation becomes
\begin{align}
\bm{\nabla} \times \bm{B}^{\rm axion} & = -\, \dfrac{\kappa}{f_{a}} \bm{\nabla} \times \left(\mu\bm{\nabla}a\right) \, . 
\end{align}
Neglecting an integration constant, one has the back-reacted magnetic field
\begin{align}
     \bm{B}^{\rm axion} = - \dfrac{\kappa}{f_{a}}\mu\,\bm{\nabla}a \, .
\end{align}
Using the mode ``10" as an example, one has $a(t, r) = R_{10}(t, r) \cos{\theta}$ and $\bm{\nabla} a(t, r) = \frac{\partial R_{10}}{\partial r}\,\cos{\theta}\,\hat{r} - \frac{R_{10}}{r}\,\sin{\theta}\,\hat{\theta}$. So, the $\bm{B}^{\rm axion}$ has nonzero components in both $\hat{r}$ and $\hat{\theta}$ directions, with the ratio different from the one of the initial dipole magnetic field.  

Note that, one has $\bm{\nabla} \cdot \bm{B}^{\rm axion} \neq 0$ and contains terms proportional to $\cos{\theta}$. Integrating over a large spherical surface, one has a zero magnetic flux, meaning the system has no net magnetic monopole. The term in the Lagrangian proportional to $\mu\,\bm{\nabla} a \cdot \bm{B}$ can be thought as a ``ferromagnetism" of the axion hair responding to the external magnetic field~\cite{Son:2004tq,Son:2007ny}. Because of the external dipole magnetic field, the magnetization direction in the north pole is opposite to the one in the south pole, which leads to no overall magnetic flux. 

The ratio of the back-reacted magnetic field on the surface over the original NS dipole magnetic field is given by
\beqa
\dfrac{|\bm{B}^{\rm axion}|}{|\bm{B}^{\rm dipole}|} \approx \left(\frac{\kappa\,\mu_0}{f_a}\right)^2 ~,
\eeqa
which is suppressed for $\mu_0 \ll f_a$. 

Similarly, the axion hair generates a new electric field proportional to $g_{a\gamma \gamma}$. Solving \eqref{eq:E-axion}, one has the back-reacted electric field as
\beqa
\bm{E}^{\rm axion} = - g_{a\gamma\gamma}\,a\,\bm{B} = - \frac{2\,\eta}{f_a}\,a\,\bm{B} ~,
\eeqa
which is along the direction of the NS dipole magnetic field. Compared to the NS rotation generated electric field $|\bm{E}^{\rm dipole}| \sim B_0 \,\Omega\,R$, the ratio is of order 
\beqa
\dfrac{|\bm{E}^{\rm axion}|}{|\bm{E}^{\rm dipole}|} \approx \dfrac{\eta\,\kappa\,\mu_0\,B_0}{\Omega\,f_a^2}  \approx (3\times 10^{-8})\,\eta\,\kappa\, \left(\dfrac{10^{11} \text{GeV}}{f_{a}} \right)^{2} \left( \dfrac{100\,\mbox{s}^{-1}}{\Omega} \right) \left( \dfrac{B_{0}}{10^{15}\,\text{G}}\right) \left(\dfrac{\mu_{0}}{10^{-3}\,\text{GeV}} \right) ~,
\eeqa
which could be sizable for an NS with a smaller rotation frequency. We also note that the two electric fields have different polarizations, which could be used as a probe~\cite{Taverna:2015vpa,Mignani:2018nmd} for the existence of the axion hair.

\section{Discussion and conclusions} \label{sec:disc}

In summary, we have examined the phenomenological consequences of electrobaryonic axions around a neutron star. Initially neglecting the effects of neutron star rotation, the prediction is that the neutron star will exhibit a static axion hair with the peak axion field value around $B_0\,\mu_0\,R/f_a$ or $(10^6\,\mbox{GeV})^2/f_a$ for $B_0 =10^{15}\,\mbox{G}$, $\mu_0 = 10^{-3}\,\mbox{GeV}$ and $R= 10$~km. Therefore, the axion field value is significantly suppressed for a large decay constant $f_a \gg 10^6$~GeV, rendering concerns about its self-interactions unnecessary. Depending on $f_a$ and neutron star properties, the total axion mass within the axion hair could surpass the total dark matter mass near the neutron star. Overall, detecting this thin axion hair around a neutron star proves to be challenging. We have also estimated the back-reacted magnetic and electric fields sourced from the axion hair. The back-reacted electric field is expected to be $10^{-8}$ times smaller than the existing electric fields, while the back-reacted magnetic field is even more negligible.

Considering the rotation effects of the NS, the anomalous interaction introduces a time-dependent source that generates the axion field. When the axion mass is below the NS rotation frequency $\Omega = 100\,\mbox{s}^{-1} \approx 7\times 10^{-14}$~eV, axion particles can be generated and propagate away from the NS. We have determined that the radiation power could be substantial [see \eqref{eq:radiation-power}], significantly contributing to the NS cooling rate.

In our study, we have incorporated the effects of general relativity on the axion hair. The impact of general relativity is insignificant for the static solution, as evident from the right panel of Fig.~\ref{fig:examplespherical}. The source can excite bound states around the NS when coupled with rotation or time-dependent effects. Treating the NS as a spherical object with constant energy density, we calculate and present the axion bound state energy levels in Fig.~\ref{fig:eigenvalues}. When the energy of the bound state matches the NS rotation frequency, resonant generation of the axion field occurs. However, for an NS with $\Omega\,R \ll 1$, we have observed that the resonantly generated axion field is subject to an additional geometrical suppression factor.

Beyond employing astrophysical objects to search for electrobaryonic axions, one could devise laboratory-based experiments to investigate the interaction described in \eqref{eq:couplings-velocity}. For example, the spatial variation of axion-like particles in the dark matter halo might induce a magnetic field in a baryon-dense environment. Additional matter-velocity suppressed interactions in \eqref{eq:couplings-velocity} could also be utilized to design experiments aimed at measuring the generated magnetic or electric fields.

Last but not least, we want to emphasize that studying additional anomalous interactions for axion-like particles, as listed in \eqref{eq:more-anomalies}, is also of great interest. This encompasses the anomaly of $U(1)_{a}U(1)_{L}[U(1)_{\rm em}]$. One could employ the lepton number current and electromagnetic field either to source the axion-like particle or to detect its presence in the dark matter halo.

\subsubsection*{Acknowledgments}
We thank Christopher T. Hill and Mrunal Korwar for the useful discussions. The work of YB is supported by the U.S. Department of Energy under the contract DE-SC-0017647. The work of CHL is supported in part by the Natural Sciences and Engineering Research Council of Canada (NSERC), TRIUMF, and the 2023 IPP Early Career Theory Fellowship. TRIUMF receives federal
funding via a contribution agreement with the National
Research Council (NRC) of Canada.

\appendix

\section{One-dimensional time-dependent solution with an infinite square well}
\label{app:infinite}

In this Appendix, we study the one-dimensional resonance effect inside the infinite well. This system approximately describes the expected effects from the gravitational well around the neutron star. The equation of motion stays the same as in Eq.~\eqref{eq:eom1} and is
\beqa
\ddot{a} - \partial_z^2 a + m_a^2 \, a  = \tilde{\kappa}\,\cos{(\Omega\,t)} \big[ \delta(z+z_{0}) - \delta(z-z_{0})\big] \, ,
\eeqa
but with the field confined in the region of $z \in [-L, L]$. Here, we want to show that the solution of this equation cannot arbitrarily grow with time, even if the source is kept on for infinite time. We also show that the only possibility of growth occurs because of the resonance effect, which becomes stronger because of the discrete nature of the energy levels.

\begin{figure}[tbh!]
\centering
    \includegraphics[width=0.7 \textwidth]{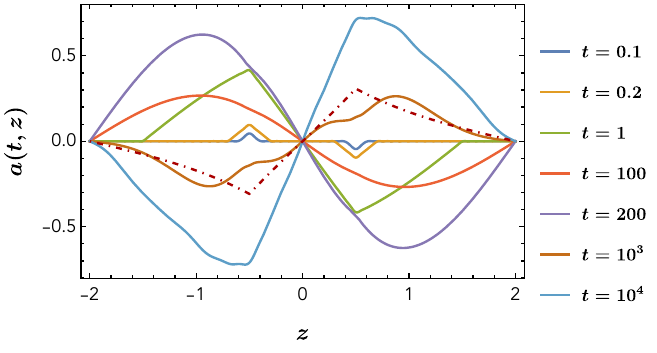}
    \caption{One example of the solution based on Eq.~\eqref{eq:infinitewell-solution} using $n_{\rm max} = 100$. The dot-dashed darker red line is the quasi-static solution. The model parameters are $L=2$, $z_0 = 1/2$, $\tilde{\kappa} = 1$, $m_a = 0.2$, $\Omega = 1$. }
    \label{fig:infinitewell}
\end{figure}

For the field confined in this infinite square well, one has the Dirichlet boundary conditions $a(t, -L) = a(t, L) = 0$. To analytically solve the equation of motion, we perform a discrete Fourier transformation with
\beqa
a(t, z) = \sum_{n = 1}^{\infty} a_n(t) \,\sin{\left[ \frac{n\,\pi\,(z + L)}{2\,L} \right]} ~.
\eeqa
The equation of motion for each mode $a_n(t)$ is
\beqa
&&\ddot{a}_n(t) + \left[ \left( \dfrac{n\,\pi}{2\,L} \right)^2 + m_a^2 \right]\,a_n(t) = \,\cos{(\Omega\,t)}\,S_n ~, \\
&&\qquad\qquad \mbox{with} \quad S_n= - \frac{2\,\tilde{\kappa}}{L}\,\cos{\left(\frac{n\,\pi}{2}\right)}\,\sin{\left(\frac{n\,\pi\,z_0}{2\,L}\right)}~,  \nonumber 
\eeqa
which vanishes for odd $n$ or parity-even wave functions. 
Subjecting to the initial conditions $a_n(t=0) = \dot{a}_n(t=0) = 0$, the solution is
\beqa
\label{eq:infinitewell-solution}
a_n(t) = S_n\,\times \, \frac{\cos{(\Omega\,t)} - \cos{\left(\sqrt{\left(\dfrac{n\,\pi}{2\,L} \right)^2 + m_a^2}\,t\right)}}{\left(\dfrac{n\,\pi}{2\,L} \right)^2 + m_a^2 - \Omega^2} ~.
\eeqa
At the initial time with $t \ll 1/\Omega, 1/m_a$, one can Taylor-expand the above solution to obtain an $t^2$ growing behavior. However, the growth stops at later times. We show one example of time-dependent solutions in Fig.~\ref{fig:infinitewell}. 

Note that there is a resonance enhancement effect when $\Omega$ and $m_a$ satisfy the relation of $\Omega^2 - m_a^2 - n_{\rm R}^2\pi^2/(4 L^2) \ll 1$ with even  $n_{\rm R}$. When this happens and defining $\epsilon\equiv [\Omega - \sqrt{m_a^2 + n_{\rm R}^2\pi^2/(4 L^2)}]/\Omega$, one can rewrite \eqref{eq:infinitewell-solution} as
\beqa
a_n(t) \approx \delta_{n, n_{\rm R}}\,S_n\,\times \, \frac{\sin{(\epsilon\,\Omega\,t)}\,\sin{(2\,\Omega\,t)}}{4\,\epsilon\,\Omega^2} ~,
\eeqa
which is enhanced by a factor of $1/\epsilon$. In the case of perfect resonance, we can have linear growth in time, but this effect will be cut short by the slowdown effect expected for $\Omega$. 

\section{Including general-relativity (GR) effects}
\label{app:GR}

In this Appendix, we explore the GR effects generated from the spacetime deformation around the neutron star. We use a simple solution to the Tolman–Oppenheimer–Volkoff (TOV) equation~\cite{Tolman:1939jz,Oppenheimer:1939ne} with a constant energy density to analytically model the neutron star. Following the notation in Ref.~\cite{Oppenheimer:1939ne}, the spherical symmetric and static line element has
\beqa
ds^2 = e^{\nu(r)}\,dt^2 - e^{\lambda(r)}\,dr^2\, - \,r^2\,\left(d\theta^2 + \sin^2{\theta}\,d\phi^2\right)~.
\eeqa
The energy momentum tensor is $T_\mu^\nu = \mbox{diag}(\rho, -p, -p, -p)$. The Einstein's equations are 
\beqa
8\pi\,G_{N}\,p &=& e^{-\lambda} \left( \frac{v'}{r} + \frac{1}{r^2}\right) - \frac{1}{r^2} ~, \\
8\pi\,G_{N}\,\rho &=& e^{-\lambda} \left( \frac{\lambda'}{r} - \frac{1}{r^2}\right) + \frac{1}{r^2} = \frac{1}{r^2}\left[1 - \frac{d}{dr}(r e^{-\lambda}))\right] ~, \\
p' &=& - \frac{(p+\rho)}{2}\,v' ~. \label{eq:TOV-3}
\eeqa
Here, prime denotes the derivative in $r$. The second equation can be integrated easily to obtain
\beqa
-g_{rr}^{-1} \equiv e^{-\lambda(r)} = \left\{
\begin{array}{cc}
 1 \,-\,\dfrac{2\,G_{N}\,m(r)}{r} & \mbox{for}~ r \le R ~, \vspace{3mm}\\
  1 \,-\,\dfrac{2\,G_{N}\,M}{r} & \mbox{for}~ r > R ~,
\end{array}  \right.
\eeqa
with the mass function $m(r) \equiv 4\pi\,\int d\tilde{r}\,\tilde{r}^2\,\rho(\tilde{r})$. The first and third equations can be combined to derive the TOV equation as
\beqa
\frac{dp}{dr} = - \frac{G_{N}}{r^2} (\rho + p) (m + 4\pi\,r^3\,p)\left(1 - \frac{2 \, G_{N}\,m}{r}\right)^{-1} ~.
\eeqa

For a constant density $\rho = \rho_0$ when $r \le R$ and zero when $r > R$, one has the analytic solution for the pressure using the boundary condition $p(R) = 0$ as
\beqa
p(r) = \rho_0 \, \frac{ \left(1- \dfrac{2 G_{N} M}{R} \right)^{1/2} - \left( 1 - \dfrac{2 G_{N} M r^2}{R^3}\right)^{1/2} }{ \left( 1 - \dfrac{2 G_{N} M r^2}{R^3}\right)^{1/2} - 3\, \left(1- \dfrac{2 G_{N} M}{R} \right)^{1/2}  } ~.
\eeqa
Here, the neutron star mass $M = 4\pi\,\rho_0\,R^3/3$. Substituting this solution back into \eqref{eq:TOV-3}, we can solve for $v(r)$ with the boundary condition of $e^{v(R)} = 1 - 2 G_{N} M/ R$. The solution is
\beqa
g_{tt} \equiv e^{v(r)} = \left\{
\begin{array}{cc}
 \dfrac{1}{4}\, \left( 3 \sqrt{1 - \dfrac{2\,G_{N}\,M}{R}}- \sqrt{ 1 - \dfrac{2\,G_{N}\,M\,r^2}{R^3} }  \right)^2 & \mbox{for}~ r \le R \,, \vspace{3mm}\\
 1 - \dfrac{2\,G_{N}\,M}{r} & \mbox{for}~ r > R \, .
 \end{array}  \right. 
\eeqa

The equation of motion of a free scalar field in the curved spacetime is
\beqa
g_{tt}^{-1}\,\partial_t^2 \,a\, + \, |g|^{-1/2}\, \partial_r \left( |g|^{1/2}\,g_{rr}^{-1}\, \partial_r a\right) - \frac{\partial_\theta(\sin{\theta} \partial_\theta a) }{r^2\,\sin \theta} - \frac{\partial^2_\phi\,a }{r^2\,\sin^2{\theta}}= - m_a^2 a  ~.
\eeqa
Here, $|g| = |\mbox{det}(g_{\mu\nu})| = | g_{tt} g_{rr} r^4 \sin^2{\theta}|$. For the region $r > R$, one has $|g| = r^4 \sin^2{\theta}$. The radial function for the $l =1$ partial wave has
\beqa
&&\frac{1}{\left(1 - \dfrac{2\,G_{N}\,M}{r}\right)}\,\partial_t^2 R_{1\mp1} - \frac{1}{r^2}\, \partial_r\left[ r^2 \left(1 - \frac{2 G_{N} M}{r}\right)\partial_r R_{1\mp1} \right] + \left(\frac{2}{r^2} + m_a^2\right) \, R_{1\mp1} \nonumber \\ 
&& \hspace{11cm} = \,\tilde{\kappa} \,e^{\pm i \Omega t}\,\delta(r -R)~.
\eeqa
For the ansatz solution, $R_{1\mp1}  = e^{\pm i \Omega t} \widetilde{R}(r)$, one then has
\beqa
- \frac{1}{r^2}\, \partial_r\left[ r^2 \left(1 - \frac{2 G_{N} M}{r}\right)\partial_r \widetilde{R} \right] +  \left(\frac{2}{r^2} + m_a^2 - \frac{\Omega^2}{\left(1 - \dfrac{2\,G_{N}\,M}{r}\right)} \right) \widetilde{R} = \tilde{\kappa} \, \delta(r - R) ~.
\eeqa
%

\setlength{\bibsep}{6pt}
\bibliographystyle{JHEP}
\bibliography{bibAXION.bib}

\end{document}